\ifx\undefined\documentclass
\documentstyle[12pt,a4,axodraw]{article}
\newcommand\mathrm[1]{{\rm #1}}
\else
\documentclass[12pt,a4paper]{article}
\usepackage{axodraw}
\fi
\newcommand{\ord}[1]{{\cal O}\left(#1\right)}
\newcommand{\sla}[1]{/\!\!\!#1}
\begin{document}
\thispagestyle{empty}

%  #] preamble:
%  #[ title page :

\begin{flushright}
INLO-PUB-8/95
\end{flushright}

\vspace{\fill}

\begin{center}
{\Large {\bf Stable Calculations for Unstable Particles:
              Restoring Gauge Invariance}}\\

\vspace{\baselineskip}
Ernestos N.~Argyres\\
{\it Institute of Nuclear Physics, NCSR `Democritos', Athens, Greece}\\[1mm]
Wim~Beenakker,
Geert Jan van Oldenborgh\\
{\it Instituut-Lorentz, Rijksuniversiteit Leiden, Netherlands}\\[1mm]
Ansgar~Denner\\
{\it Institut f\"ur Theoretische Physik, Universit\"at W\"urzburg,
Germany}\\[1mm]
Stefan~Dittmaier\\
{\it Institut f\"ur Physik, Universit\"at Bielefeld, Germany}\\[1mm]
Jiri~Hoogland, Ronald~Kleiss\\
{\it NIKHEF-H, Amsterdam, Netherlands}\\[1mm]
Costas G.~Papadopoulos\\
{\it Physics Department, University of Durham, UK}\\
{\it CERN, Theory Division, Geneva, Switzerland}\\[1mm]
Giampiero~Passarino\\
{\it Dipartimento di Fisica Teorica, Universit\`a di Torino, Italy}\\
{\it INFN, Sezione di Torino, Italy}\\
\vspace{2\baselineskip}
{\bf Abstract}
\end{center}

We discuss theoretical and phenomenological aspects of the use
of boson propagators with energy-dependent widths in predictions for
high-energy scattering processes.
In general, gauge invariance is violated
in such calculations. We discuss several approaches to restore gauge
invariance, necessary for a reliable result. The most promising method is
the addition of the relevant parts of the fermionic corrections,
which fulfills all Ward identities.
The numerical difference between this and other approaches is studied.
A number of recommendations are given for LEP2 computations.
%\tableofcontents

\vspace{\fill}
\newpage
\pagestyle{plain}
\setcounter{page}{1}

%  #] title page :
%  #[ introduction :

\section{Introduction}
Monte Carlo calculations for scattering processes for LEP2
and higher-energy colliders are required to have a precision
of better than one percent.
It is obvious that under such circumstances the assumption that the
$W$ and $Z$ vector bosons are stable particles, produced on-shell,
is a gross misrepresentation of the physics.  Rather, one has
to describe them as resonances, with a finite width so as to avoid
singularities inside the physical phase space.
In field theory, such widths arise naturally from the
imaginary parts of higher-order
diagrams describing the boson self-energies, resummed to all orders.
This procedure has been used with great success in the past: indeed,
the $Z^0$ resonance can be described to very high numerical accuracy.
However, in doing a Dyson summation
of self-energy graphs, we are singling out only a very limited subset
of all the possible higher-order diagrams. It is therefore not
surprising that one often ends up with a result that retains
some gauge dependence.

In itself, this is not necessarily a problem if one treads wearily.
An example is the situation at LEP1.  Here, a careful separation of the
gauge-invariant subsets can be performed, leading to a result which has no
significant gauge dependence.
For processes that become important at LEP2, the situation is in several
cases more complicated. Since gauge invariance is intimately connected with
the high-energy behaviour of the theory, it is to be expected that
the effects of gauge violation become worse if the scattering process
under study contains a ratio of masses, or of momentum transfers, that
becomes large. An example, which we shall study in this paper, is
the production and hadronic decay of a single $W$ in the process
\begin{equation}
e^+e^-
\to e^-\bar{\nu}_eW^+
\to e^-\bar{\nu}_e u\bar{d}\;\;.
\label{singleWprocess}
\end{equation}
Here, the electron may
emit a virtual photon, whose $q^2$ can be as small as $m_e^2$, where
$m_e$ is the electron mass: with a total center of mass energy of $\sqrt{s}$
available,
we have a mass ratio of $s/m_e^2 = \ord{10^{11}}$, large enough
to amplify even a tiny gauge violation in a disastrous way\footnote{This was
noted already in Ref.~\cite{BerendsW}, and investigated further in
Ref.~\cite{KEKtchan}.}.  An other, currently
academic, situation, connected with SU(2) rather than U(1)$_{e.m.}$, is the
gauge cancellation which prevents the cross-section for $e^+e^-\to W^+W^-$
from blowing up for high energies.

In order to arrive at phenomenologically reliable predictions,
various approaches can be followed\footnote{A few were investigated in
Ref.~\cite{OGamma}.}. In the first place, we may try to
convince ourselves that, for the particular problem under study, the
situation is actually not so bad. For instance, this is the case in
the above-mentioned LEP1 processes. In processes like
\begin{equation}
e^+ e^- \to \gamma,Z \to \mu^+\mu^-\;\;,
\end{equation}
there is no obvious dangerous large ratio of masses at energies
around the $Z$ mass, as the relevant ratio is $s/M_{_Z}^2$.
One might therefore hope that, by the imposition of a cut on the
electron scattering angle in the process (\ref{singleWprocess}), which
effectively leads to a lower bound on the $q^2$ of the virtual photon,
the effects of gauge violation can be mitigated. This is, for instance,
an implicit assumption made in the {\tt Excalibur} Monte Carlo
\cite{ExcaliburCPC}.
Of course, such a hope has to be borne out by comparison with a
gauge-invariant calculation.

One may sidestep the problem by simply performing
the calculation of the matrix elements without any width,
and only at the end use some ad-hoc prescription like the
following \cite{Zeppenfeld&Co,KEKtchan}. Let the mass and width of a boson be
given by $M$ and $\Gamma$, respectively, and its momentum by $q^{\mu}$
($\Gamma$ may depend on $q^2$). Then, if we multiply the matrix element by
$(q^2-M^2)/(q^2-M^2+iM\Gamma)$, the pole at $q^2=M^2$ is
softened into a resonance, at the expense of mistreating the non-resonant
parts.
It should be noted, that there are examples, where this `fudge-factor
scheme' leads to deviations up to 30\% \cite{ZeppenfeldBaur}.

Another way to sidestep the problem is to use the `fixed-width scheme',
i.e., to systematically replace $1/(q^2-M^2)$ by $1/(q^2-M^2+iM\Gamma)$, also
for $q^2<0$.
This gives U(1)$_{e.m.}$-current conservation, but it has no physical
motivation.
In perturbation theory the propagator for space-like momenta does not develop
an imaginary part. Moreover, the fixed-width approximation violates
the SU(2)$\times$U(1) Ward identities.  Note, however, that this does not lead
to a bad high-energy behaviour in $e^+e^-\to$ 4 fermions, as the unitarity
cancellations do not involve the masses of the $W$ and $Z$ bosons.  In the
case of $e^+e^-\to$ 6 fermions (e.g., $W_L$ scattering) the occurrence of
$W$-mass dependent couplings means the unitarity cancellations are violated by
a fixed width.

A minimalist's approach is to make use of the fact that the residue of the
amplitude at the (complex) pole is gauge-invariant
\cite{VeltmanUnstable,Stuart1}.
One can split the amplitude accordingly, and resum only this pole.  In this way
higher-order corrections can be included consistently
\cite{Andre&Geert&Daniel}.  However, this `pole scheme' breaks down near
thresholds, and has problems with the radiation of photons of energy $E_\gamma
\approx \Gamma$.

Finally, one may determine the minimal set of
Feynman diagrams that is necessary to compensate for the gauge
violation caused by the self-energy graphs, and try to include these.
This is obviously the theoretically most satisfying solution, but
it may cause an increase in the complexity of the matrix elements
and a consequent slowing down of the numerical calculations.
For the vector bosons, the lowest-order widths are given by the imaginary
parts of the fermion loops in the one-loop self-energies.  It is therefore
natural to include the other possible fermionic one-loop corrections
\cite{Simma,BeenakkerResum}.
These fermionic contributions form a gauge-independent subset
and obey all Ward identities exactly, even with resummed propagators
\cite{LongPub}.
This implies that the high-energy and collinear limits are properly behaved.
In contrast to all other schemes mentioned above, the scheme
proposed here does not modify the theory by hand but selects
an appropriate set of higher-order contributions to restore gauge invariance.

To solve the problem of gauge invariance related to the width,
we only have to consider here the imaginary parts of these diagrams%
\footnote{As the Ward identities are linear, we can separate the real and
imaginary parts.}.
This scheme should work properly for all tree-level
calculations involving resonant W-bosons and Z-bosons or other
particles decaying exclusively into fermions. For resonating
particles decaying also into bosons, such as the top quark, gauge
independence is lost. For simplicity, we take all fermions in
loops to be massless in the following.

The justification, including masses and the details of the proper
resummation and renormalization procedure, will be given in a later
publication \cite{LongPub}.
The method has already been mentioned, and implemented in the $s$ and $t$
channel as a Monte-Carlo generator for the processes $e^+e^-\to \ell\nu_\ell q
\bar{q}'$, in Ref.~\cite{CostasWWV}.
For the special case of $q\bar{q}'\to\ell\nu_\ell\gamma$,
this approach was also used by Baur and Zeppenfeld \cite{ZeppenfeldBaur},
who found that electromagnetic current conservation was restored by fermion
loops that essentially rescale the $WW\gamma$ vertex.

Although the proposed scheme is well-justified in standard perturbation
theory, it should be stressed that all reparation schemes are arbitrary to
a greater or lesser extent: since the Dyson summation must necessarily
be taken to all orders of perturbation theory, and we are not able
to compute the complete set of {\em all\/} Feynman diagrams to {\em all\/}
orders, the various schemes differ even
if they lead to formally gauge-invariant results.
It is then a numerical question how much their predictions differ.\\

The outline of this paper is as follows. In the next section,
we study the process of Eq.~\ref{singleWprocess}, with emphasis on
its small-angle behaviour. We show how gauge invariance gets violated
by the imposition of an energy-dependent width, leading to
completely wrong results.  This is repaired by the inclusion of fermionic
corrections to the three-boson vertex.  The electromagnetic current is
conserved again, and all Ward identities are satisfied.
We discuss the connection between our result and that of
Ref.~\cite{ZeppenfeldBaur}.  In section \ref{sec:numerics}, we present
numerical comparisons between the various
reparation schemes for the process (\ref{singleWprocess}).
We finish with a number of conclusions and recommendations.

%  #] introduction :
%  #[ gauge cancellation 1 :

\section{Gauge cancellations in $e^-e^+\to e^-\bar{\nu}_e u \bar{d}$}

\begin{figure}[htb]
\begin{center}
\begin{picture}(90,100)(0,0)
\ArrowLine(10,90)(35,80)
\ArrowLine(35,80)(80,90)
\ArrowLine(80,10)(35,20)
\ArrowLine(35,20)(10,10)
\ArrowLine(85,35)(70,50)
\ArrowLine(70,50)(85,65)
\Photon(35,80)(40,50){2}{4}
\Photon(40,50)(35,20){2}{4}
\Photon(40,50)(70,50){2}{4}
\Vertex(35,80){1.2}
\Vertex(35,20){1.2}
\Vertex(40,50){1.2}
\Vertex(70,50){1.2}
\put(08,90){\makebox(0,0)[r]{$e^-$}}
\put(08,10){\makebox(0,0)[r]{$e^+$}}
\put(82,90){\makebox(0,0)[l]{$e^-$}}
\put(82,10){\makebox(0,0)[l]{$\bar{\nu}_e$}}
\put(87,65){\makebox(0,0)[l]{$u$}}
\put(87,35){\makebox(0,0)[l]{$\bar{d}$}}
\put(34,65){\makebox(0,0)[r]{$\gamma$}}
\put(35,35){\makebox(0,0)[r]{$W$}}
\put(57,55){\makebox(0,0)[b]{$W^+$}}
\end{picture}
\qquad
\begin{picture}(90,90)(0,-10)
\ArrowLine(10,70)(35,60)
\ArrowLine(35,60)(80,70)
\ArrowLine(80, 0)(55,18)
\ArrowLine(55,18)(35,20)
\ArrowLine(35,20)(10,10)
\ArrowLine(85,25)(70,35)
\ArrowLine(70,35)(85,50)
\Photon(35,60)(35,20){2}{5}
\Photon(55,18)(70,35){2}{3}
\Vertex(35,60){1.2}
\Vertex(35,20){1.2}
\Vertex(55,18){1.2}
\Vertex(70,35){1.2}
\put(08,70){\makebox(0,0)[r]{$e^-$}}
\put(08,10){\makebox(0,0)[r]{$e^+$}}
\put(82,70){\makebox(0,0)[l]{$e^-$}}
\put(82, 0){\makebox(0,0)[l]{$\bar{\nu}_e$}}
\put(87,50){\makebox(0,0)[l]{$u$}}
\put(87,25){\makebox(0,0)[l]{$\bar{d}$}}
\put(30,40){\makebox(0,0)[r]{$\gamma$}}
\put(70,30){\makebox(0,0)[br]{$W^+$}}
\end{picture}
\\
\begin{picture}(90,120)(0,0)
\ArrowLine(10,110)(40,100)
\ArrowLine(40,100)(80,110)
\ArrowLine(80,10)(40,20)
\ArrowLine(40,20)(10,10)
\ArrowLine(85,40)(50,45)
\ArrowLine(50,45)(50,75)
\ArrowLine(50,75)(85,80)
\Photon(40,100)(50,75){2}{4}
\Photon(40, 20)(50,45){2}{4}
\Vertex(40,100){1.2}
\Vertex(40, 20){1.2}
\Vertex(50,45){1.2}
\Vertex(50,75){1.2}
\put(08,110){\makebox(0,0)[r]{$e^-$}}
\put(08,10){\makebox(0,0)[r]{$e^+$}}
\put(82,110){\makebox(0,0)[l]{$e^-$}}
\put(82,10){\makebox(0,0)[l]{$\bar{\nu}_e$}}
\put(87,80){\makebox(0,0)[l]{$u$}}
\put(87,40){\makebox(0,0)[l]{$\bar{d}$}}
\put(41,85){\makebox(0,0)[r]{$\gamma$}}
\put(42,35){\makebox(0,0)[r]{$W$}}
\end{picture}
\qquad
\begin{picture}(90,120)(0,0)
\ArrowLine(10,110)(40,100)
\ArrowLine(40,100)(80,110)
\ArrowLine(80,10)(40,20)
\ArrowLine(40,20)(10,10)
\ArrowLine(85,40)(65,60)
\Line(65,60)(50,75)
\ArrowLine(50,75)(50,45)
\Line(50,45)(65,60)
\ArrowLine(65,60)(85,80)
\Photon(40,100)(50,75){2}{4}
\Photon(40, 20)(50,45){2}{4}
\Vertex(40,100){1.2}
\Vertex(40, 20){1.2}
\Vertex(50,45){1.2}
\Vertex(50,75){1.2}
\put(08,110){\makebox(0,0)[r]{$e^-$}}
\put(08,10){\makebox(0,0)[r]{$e^+$}}
\put(82,110){\makebox(0,0)[l]{$e^-$}}
\put(82,10){\makebox(0,0)[l]{$\bar{\nu}_e$}}
\put(87,80){\makebox(0,0)[l]{$u$}}
\put(87,40){\makebox(0,0)[l]{$\bar{d}$}}
\put(41,85){\makebox(0,0)[r]{$\gamma$}}
\put(42,35){\makebox(0,0)[r]{$W$}}
\end{picture}
\end{center}
\caption[]{The four diagrams of the process $e^-e^+\to e^-\bar{\nu}_e u
\bar{d}$ which are considered here.}
\label{diagrams_eeenuud}
\end{figure}
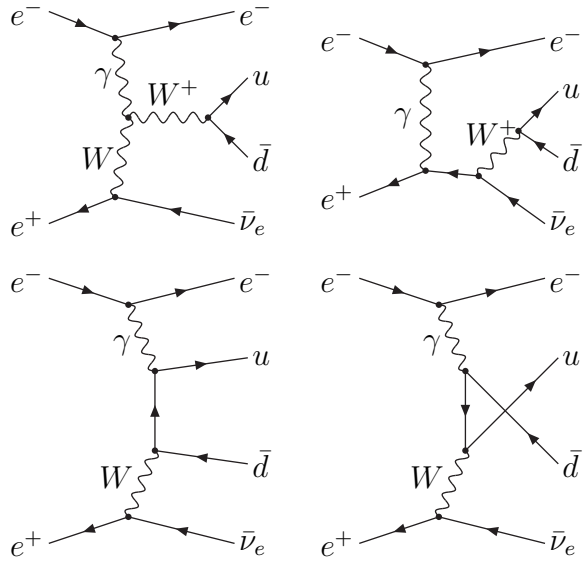

In this section, we consider the process
\begin{equation}
\label{process1}
e^-(p_1)\;e^+(k_1)\;\;\to\;\;e^-(p_2)\;\bar{\nu}_e(k_2)\;u(p_u)\;
\bar{d}(p_d)\;\;,
\end{equation}
and especially concentrate on small scattering angles $\theta$ for the
electron.
We keep the mass of the electron finite, but shall neglect all other
fermion masses (also that of the positron), so that we shall not have to worry
about diagrams with Higgs ghosts connected to the positron or quark lines.
The massive case can be treated analogously; this will be
covered in Ref.~\cite{LongPub}.
Under these assumptions, we have to consider the subset of
four Feynman diagrams given in Fig.~\ref{diagrams_eeenuud}, which conserves
the electromagnetic current.

%  #[ diagrams :

%  #] diagrams :

The  matrix element ${\cal M}$ is given by
\begin{eqnarray}
{\cal M}
  & = &
  {\cal M}^{\mu}\;
  J_\mu\;\;,\;\;\;
  {\cal M}^{\mu}
  =
  \sum_{i=1}^4{\cal M}_i^{\mu}\;\;,\nonumber\\
{\cal M}_1^{\mu}
  & = &
  Q_{_W}\;
  P_{_W}(p_+^2)\;P_{_W}(p_-^2)\;
  V^{\alpha\beta\mu}(p_+,-p_-,-q)  {\cal M}^0_{\alpha\beta}\;\;,\nonumber\\
{\cal M}_2^{\mu}
  & = &
  \hphantom{-}
  4iQ_e g_w^2
  \;P_{_W}(p_+^2)\;
  \bar{v}_-(k_1)\gamma^\mu{\sla{k}_1+\sla{q}\over(k_1+q)^2}
     \gamma^\beta v_-(k_2)\;\;
  \bar{u}_-(p_u)\gamma_\beta v_-(p_d)\;\;,\nonumber\\
{\cal M}_3^{\mu}
  & = &
  -4iQ_u g_w^2
  \;P_{_W}(p_-^2)\;
  \bar{u}_-(p_u)\gamma^\mu{\sla{p}_u-\sla{q}\over(p_u-q)^2}
    \gamma^\beta v_-(p_d)\;\;
     \bar{v}_-(k_1)\gamma_\beta v_-(k_2)\;\;,\nonumber\\
{\cal M}_4^{\mu}
  & = &
  -4iQ_d g_w^2
  \;P_{_W}(p_-^2)\;
%%\nonumber\\
%%  & &
  \bar{u}_-(p_u)\gamma^\beta{\sla{q}-\sla{p}_d\over(p_d-q)^2}
    \gamma^\mu v_-(p_d)\;\;
  \bar{v}_-(k_1)\gamma_\beta v_-(k_2)\;\;,\nonumber\\
{\cal M}^0_{\alpha\beta}
  & \equiv &
  4i g_w^2\;
  \bar{v}_-(k_1)\gamma_\beta v_-(k_2)\;\;
  \bar{u}_-(p_u)\gamma_\alpha v_-(p_d)\;\;.
\label{fourdiagrams}
\end{eqnarray}
The spinors are written in a compact form,
$u_-(p) \equiv {1\over2}(1-\gamma^5) u(p)$ and
\begin{eqnarray}
& & p_+ = p_u+p_d\;\;,\;\;p_- = k_1-k_2\;\;,\;\;q = p_1-p_2\;\;,\nonumber\\
& & \left[P_{_W}(s)\right]^{-1} = s-M_{_W}^2+i\gamma_{_W}(s)\;\;,
\end{eqnarray}
where $M_{_W}$ is the $W$ mass and
$\gamma_{_W}$ denotes the imaginary part of the inverse $W$ propagator,
which for the moment is introduced as a purely phenomenological
device in order to avoid the singularities.
The charged weak coupling constant $g_w$ is given by
$g_w^2 = M_{_W}^2G_F/\sqrt{2}$,
$Q_i$ is the electric charge of particle $i$, and
\begin{equation}
V^{\mu_1\mu_2\mu_3}(p_1,p_2,p_3) = (p_1-p_2)^{\mu_3}g^{\mu_1\mu_2} +
(p_2-p_3)^{\mu_1}g^{\mu_2\mu_3} + (p_3-p_1)^{\mu_2}g^{\mu_3\mu_1}\;\;.
\end{equation}
If we use conservation of the charged current in the massless fermion
lines, we may write
\begin{equation}
V^{\alpha\beta\mu}(p_+,-p_-,-q) =
 (2p_+-q)^\mu g^{\alpha\beta} + 2q^\alpha g^{\beta\mu} - 2p_+^\beta
    g^{\mu\alpha}\;\;.
\end{equation}
The photon source is given by
\begin{equation}
J^{\mu} = {Q_e \over q^2} \bar{u}(p_2)\gamma^{\mu} u(p_1)\;\;.
\end{equation}
Note that the electrons can have two spin states each, but the
massless fermions only contribute when they are left-handed.

The matrix element, squared and averaged over the spins of the
incoming fermions, reads
\begin{eqnarray}
\left\langle|{\cal M}|^2\right\rangle & = &
H^{\mu\nu}{\cal M}_\mu\overline{{\cal M}}_\nu\;\;,\nonumber\\
H^{\mu\nu} & = & {1\over4}\sum_{\mbox{\tiny spins}}J^\mu\bar{J}^\nu\;\;
= \;\; {Q_e^2 \over q^4}
  \left[ p_1^\mu p_2^\nu + p_1^\nu p_2^\mu + {q^2\over2}g^{\mu\nu} \right]\;\;.
\end{eqnarray}
Note the occurrence of $q^{-4}$: we may approximate
\begin{equation}
|q^2| \sim {m_e^2 R^4\over S(S-R^2)} + {S-R^2\over2}(1-\cos\theta)\;\;,
\end{equation}
where $\sqrt{S}$ is the total energy, $m_e$ the electron mass, $\theta$
the electron scattering angle, and $R^2=(p_u+p_d+k_2)^2$: therefore,
$|q^2|$ can be as small as $\ord{m_e^2}$.
Let us now consider the numerical behaviour of $H^{\mu\nu}$. Using
the old approach of Ref.~\cite{Gutbrod&Rek}, we define
\begin{equation}
r^\mu = p_1^\mu - \beta p_2^\mu\;\;\;,\;\;\;\beta = p_1^0/p_2^0\;\;,
\end{equation}
so that $r^0=0$ and $(r)^2 = (1-\beta)^2m_e^2 + \beta q^2$; therefore,
as $|q^2|$ becomes small, each individual component of $r^\mu$
also becomes small. We may now write
\begin{equation}
H^{\mu\nu}\! = {Q_e^2 \over q^4(1-\beta)^2}\left[
2r^\mu r^\nu +2\beta q^\mu q^\nu - (1\!+\!\beta)(r^\mu q^\nu+r^\nu q^\mu)
+ {1\over2}(1\!-\!\beta)^2q^2g^{\mu\nu} \right].
\end{equation}
The unwanted $q^{-4}$ behaviour of the cross-section will be
mitigated to a $q^{-2}$ behaviour, {\em provided}
\begin{equation}
q^\mu {\cal M}_\mu = 0\;\;.
\label{currcons}
\end{equation}
This conservation of electromagnetic current is seen to be
extremely important here: any small violation of it will be
punished by a huge factor $\ord{S/m_e^2}$.
Multiplying $q^\mu$ into the four diagrams of Eq.~\ref{fourdiagrams}, we obtain
\begin{eqnarray}
W & \equiv & q^\mu{\cal M}_\mu \nonumber\\
  & = & {\cal M}_0\left\{
  (p_+^2-p_-^2)Q_{_W}\;P_{_W}(p_+^2)\;P_{_W}(p_-^2)\right.\nonumber\\
& & \hphantom{{\cal M}_0}
   \left.\mbox{}
   +Q_e\;P_{_W}(p_+^2) - \left( Q_d-Q_u\right)\;
     P_{_W}(p_-^2) \right\}\;\;,\nonumber\\
{\cal M}_0
  & \equiv &
  {\cal M}^0_{\alpha\beta}\,g^{\alpha\beta}\;\;.
\end{eqnarray}
By taking $\gamma_{_W}=0$, and considering the two poles at
$p_+^2=M_{_W}^2$ and at $p_-^2=M_{_W}^2$, we get from the condition
Eq.~\ref{currcons} that
\begin{equation}
Q_{_W} = Q_e = Q_d-Q_u\;\;,
\end{equation}
the obvious condition of charge conservation. Therefore, we have
\begin{equation}
W = -i\;Q_e{\cal M}_0\;P_{_W}(p_+^2)\;P_{_W}(p_-^2)\;
        \left(\gamma_{_W}(p_+^2)-\gamma_{_W}(p_-^2)\right)\;\;.
\label{gaugecancellation1}
\end{equation}
Current conservation is therefore violated unless
$\gamma_{_W}(p_+^2)=\gamma_{_W}(p_-^2)$. The most naive treatment of a
Breit-Wigner resonance uses a {\em fixed width\/} approximation, with
\begin{equation}
\gamma_{_W}(s)_{\mbox{\tiny fixed width}} = M_{_W}\Gamma_{_W}\;\;.
\end{equation}
The nominal width of an on-shell $W$ is given by
\begin{equation}
\label{widthw}
\Gamma_{_W} = \sum_{\mbox{\tiny doublets}}\;N_f\;
        {G_FM_{_W}^3\over6\pi\sqrt{2}}\;\;,
\end{equation}
involving a sum over all massless fermion doublets with $N_f$ (=1 or 3)
colours.   In this approximation, there is evidently no violation of
electromagnetic current conservation.

The difficulty with the fixed-width approximation is that it cannot
be justified from field theory. Indeed, in field theory the
propagator only develops a complex pole off the real axis if we
perform a Dyson summation of the self-energy graphs to all orders.
This self-energy is inherently energy-dependent: to a good
approximation\footnote{Eq.~\ref{eq:Wwidth} exactly takes into account the
contributions of massless fermions, but it should be noted that above the $W$
mass there is a contribution from the $W\gamma$ self-energy diagram,
which has to be treated perturbatively.}, we may write
\begin{eqnarray}
  \gamma_{_W}(s) & = & {\Gamma_{_W}\over M_{_W}}s\;\;\;,\;\;\;s\ge0\;\;,
\nonumber\\
  \gamma_{_W}(t) & = & 0\;\;\;,\;\;\;t<0\;\;.
\label{eq:Wwidth}
\end{eqnarray}
Consequently, propagators with space-like momenta cannot
acquire finite widths in contradiction to the fixed-width scheme.

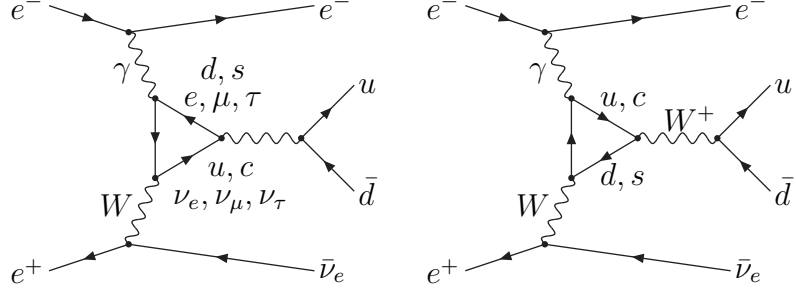
\begin{figure}[tb]
\begin{center}
\begin{picture}(130,120)(0,0)
\ArrowLine(10,110)(40,100)
\ArrowLine(40,100)(110,110)
\ArrowLine(110,10)(40,20)
\ArrowLine(40,20)(10,10)
\ArrowLine(125,40)(105,60)
\ArrowLine(105,60)(125,80)
\Photon(40,100)(50,75){2}{4}
\Photon(40, 20)(50,45){2}{4}
\ArrowLine(50,75)(50,45)
\ArrowLine(50,45)(75,60)
\ArrowLine(75,60)(50,75)
\Photon(75,60)(105,60){2}{4}
\Vertex(40,100){1.2}
\Vertex(40, 20){1.2}
\Vertex(50, 45){1.2}
\Vertex(50, 75){1.2}
\Vertex(105,60){1.2}
\Vertex( 75,60){1.2}
\put(08,110){\makebox(0,0)[r]{$e^-$}}
\put(08,10){\makebox(0,0)[r]{$e^+$}}
\put(112,110){\makebox(0,0)[l]{$e^-$}}
\put(112,10){\makebox(0,0)[l]{$\bar{\nu}_e$}}
\put(127,80){\makebox(0,0)[l]{$u$}}
\put(127,40){\makebox(0,0)[l]{$\bar{d}$}}
\put(41,85){\makebox(0,0)[r]{$\gamma$}}
\put(42,35){\makebox(0,0)[r]{$W$}}
\put(61,70){\makebox(0,0)[bl]{\shortstack{$d,s$\\$e,\mu,\tau$}}}
\put(57,51){\makebox(0,0)[tl]{\shortstack{$u,c$\\$\nu_e,\nu_\mu,\nu_\tau$}}}
\end{picture}
\qquad
\begin{picture}(130,120)(0,0)
\ArrowLine(10,110)(40,100)
\ArrowLine(40,100)(110,110)
\ArrowLine(110,10)(40,20)
\ArrowLine(40,20)(10,10)
\ArrowLine(125,40)(105,60)
\ArrowLine(105,60)(125,80)
\Photon(40,100)(50,75){2}{4}
\Photon(40, 20)(50,45){2}{4}
\ArrowLine(50,45)(50,75)
\ArrowLine(75,60)(50,45)
\ArrowLine(50,75)(75,60)
\Photon(75,60)(105,60){2}{4}
\Vertex(40,100){1.2}
\Vertex(40, 20){1.2}
\Vertex(50, 45){1.2}
\Vertex(50, 75){1.2}
\Vertex(105,60){1.2}
\Vertex( 75,60){1.2}
\put(08,110){\makebox(0,0)[r]{$e^-$}}
\put(08,10){\makebox(0,0)[r]{$e^+$}}
\put(112,110){\makebox(0,0)[l]{$e^-$}}
\put(112,10){\makebox(0,0)[l]{$\bar{\nu}_e$}}
\put(127,80){\makebox(0,0)[l]{$u$}}
\put(127,40){\makebox(0,0)[l]{$\bar{d}$}}
\put(41,85){\makebox(0,0)[r]{$\gamma$}}
\put(42,35){\makebox(0,0)[r]{$W$}}
\put(61,70){\makebox(0,0)[bl]{$u,c$}}
\put(61,51){\makebox(0,0)[tl]{$d,s$}}
\put(95,63){\makebox(0,0)[b]{$W^+$}}
\end{picture}
\end{center}
\caption[]{The extra fermionic diagrams needed to cancel the gauge-breaking
terms.}
\label{extra_diagrams_eeenuud}
\end{figure}

The theoretically most satisfying way to restore gauge-invariance seems to
be the addition of one-loop vertex-corrections, which cancel the imaginary
part in the Ward identities. In the process above, this boils down to adding
the imaginary parts of the diagrams of Fig.~\ref{extra_diagrams_eeenuud}.
These are given by

%  #[ diagrams2 :

%  #] diagrams2 :

\begin{equation}
\label{eq:M5}
{\cal M}_5^{\mu}  =  {i\over16\pi}{\cal M}^0_{\alpha\beta}\;P_{_W}(p_+^2)\;
    P_{_W}(p_-^2)\;g_w^2\sum_{\mbox{\tiny doublets}}N_f\left( Q_d-Q_u \right)\;
                Z^{\alpha\beta\mu}\;\;,
\end{equation}
where we included the appropriate
colour factor for the doublet, $N_f$.  Using cutting rules, we calculate
\begin{equation}
\label{eq:Z1}
Z^{\alpha\beta\mu} = {1\over2\pi} \int d\Omega
     \;\mbox{Tr}\left[ \sla{r}_1\gamma^\mu{\sla{r}_1-\sla{q}\over(r_1-q)^2}
                         \gamma^\beta\sla{r}_2\gamma^\alpha
                  \right]\;\;,
\end{equation}
which is the imaginary part of the triangle insertions. The momenta $r_1$ and
$r_2$ are the momenta of the cut fermion lines with $p_+=r_1+r_2$.
The expression $Z^{\alpha\beta\mu}$ satisfies the following three Ward
identities
\begin{eqnarray}
Z^{\alpha\beta\mu} q_\mu
  & = & -{8\over3}\left( p_+^\alpha p_+^\beta - p_+^2 g^{\alpha\beta}
                                         \right) \;\;,\nonumber\\
Z^{\alpha\beta\mu} p^+_\alpha
  & = & 0 \;\;,\nonumber\\
Z^{\alpha\beta\mu} p^-_\beta
  & = & +{8\over3}\left( p_+^\mu p_+^\alpha - p_+^2 g^{\mu\alpha}
                                        \right) \;\;.
\label{wiz}
\end{eqnarray}
Because of the anomaly cancellation we have no explicit contributions from the
part containing $\gamma^5$.  Possible effects due to a top quark remain
to be studied.  Attaching the photon momentum $q_\mu$ to the sum of the
diagrams ${\cal M}_5^{\mu}$ gives
\begin{eqnarray}
W_{\mathrm{add}} & \equiv & q_\mu{\cal M}_5^{\mu} \nonumber\\
  & = &
  i\;Q_e{\cal M}_0\;P_{_W}(p_+^2)\;P_{_W}(p_-^2)\;
  g_w^2\sum_{\mbox{\tiny doublets}} N_f{p_+^2\over6\pi}\nonumber\\
  & = &
  i\;Q_e{\cal M}_0\;P_{_W}(p_+^2)\;P_{_W}(p_-^2)\;
  \Gamma_{_W} {p_+^2\over M_{_W}}\nonumber\\
  & = &
  i\;Q_e{\cal M}_0\;P_{_W}(p_+^2)\;P_{_W}(p_-^2)\;
  \gamma_{_W}(p_+^2)\;\;,
\end{eqnarray}
where we used the Ward identity of Eq.~\ref{wiz}, the definition of the
nominal $W$ width, Eq.~\ref{widthw}, as well as the fact that
the external charged currents are conserved for massless fermions.
It is clear that the extra diagrams exactly cancel
the imaginary part in Eq.~\ref{gaugecancellation1}.
Hence electromagnetic current conservation is restored.

Using the fact that all external fermionic currents in process
(\ref{process1}) are conserved
one finds for the compensating correction (Eq.~\ref{eq:Z1})
\begin{eqnarray}
\label{eq:Z}
Z^{\alpha\beta\mu} & = & p_+^\mu q^\beta q^\alpha c_0
                + g^{\mu\beta}q^\alpha c_1
                + g^{\mu\alpha}q^\beta c_2
                + g^{\alpha\beta}p_+^\mu c_3 \;\;,\nonumber\\[1mm]
c_0 & = & 32\,\frac{p_+^2p_-^2q^2}{\lambda^2}\,\left\{ \left[ 10\,
          \frac{p_+^2p_-^2q^2}{\lambda}+p_+^2+p_-^2+q^2 \right]\,f_0
          \right.\nonumber\\
    &   & \hphantom{32\,\frac{p_+^2p_-^2q^2}{\lambda^2}A}
          \vphantom{32\,\frac{p_+^2p_-^2q^2}{\lambda^2}}
          \left.
          +\,20p_+^2\,\frac{p_-\!\!\cdot q}{\lambda} - \frac{8}{3}
          + \frac{2}{3}(p_-^2+q^2)\,\frac{p_-\!\!\cdot q}{p_-^2q^2} \right\}
          \;\;,\nonumber\\
c_1 & = & -8\,\frac{p_+^2p_-^2q^2}{\lambda}\,\left\{ \left[ 2p_+^2\,
          \frac{p_-\!\!\cdot q}{\lambda}-1\right]\,f_0
          + 4\,\frac{p_+^2}{\lambda}
          - 2\,\frac{p_-\!\!\cdot q}{p_-^2q^2}
          + \frac{1}{3}\,\frac{p_+^2}{p_-^2q^2}
          \right\}\;\;,\nonumber\\
c_2 & = & -8\,\frac{p_+^2p_-^2q^2}{\lambda}\,\left\{ \left[ 2p_-^2\,
          \frac{p_+\!\!\cdot q}{\lambda}+1\right]\,f_0
          + 4\,\frac{p_+\!\!\cdot p_-}{\lambda}
          + \frac{4}{3}\,\frac{p_+\!\!\cdot p_-}{p_-^2q^2} - \frac{1}{q^2}
          \right\}\;\;,\nonumber\\
c_3 & = & +8\,\frac{p_+^2p_-^2q^2}{\lambda}\,\left\{ \left[ 2q^2\,
          \frac{p_+\!\!\cdot p_-}{\lambda}+1\right]\,f_0
          + 4\,\frac{p_+\!\!\cdot q}{\lambda}
          + \frac{4}{3}\,\frac{p_+\!\!\cdot q}{p_-^2q^2} - \frac{1}{p_-^2}
          \right\}\;\;,\nonumber\\
f_0 & = & -{2\over \sqrt{\lambda}}\ln\left({2(p_-\!\!\cdot q)+\sqrt{\lambda}
          \over 2(p_-\!\!\cdot q)-\sqrt{\lambda}}\right) \;\;,\;\;
\lambda  \equiv
         4(p_-\!\!\cdot q)^2-4p_-^2q^2\;\;.
\end{eqnarray}
This expression, inserted in Eq.~\ref{eq:M5}, gives the correction to the
$WW\gamma$ vertex to be used in explicit calculations.

Now we discuss the result of Ref.~\cite{ZeppenfeldBaur}. They computed
the process  $q\bar{q}'\to\ell\nu_\ell\gamma$ with dressed
propagators for the $W$'s and an on-shell photon. Note, that in the process
of single $W$ production, one had $q+p_-=p_+$. In the following process, we
have $p_-=q+p_+$. Hence, we have put $q\to -q$ with respect to the former
definitions of the momenta. The invariant momentum squared flowing
through both $W$'s is positive and hence the running width is non-zero in both
propagators. Without addition of extra diagrams, the corresponding amplitude
will again not be gauge-invariant. Using the previous result in this section,
it is easy to see that, with $q^2=0$, one gets for the two cut diagrams
corresponding to the cut $p_+^2>0$:
\begin{eqnarray}
Z^{\alpha\beta\mu} & = & \hphantom{\mbox{}+\mbox{}} {16\over3}{p_+^2\over a}
                    \left(
                           g^{\alpha\beta}p_+^\mu
                          +g^{\mu\alpha}q^\beta
                          -g^{\mu\beta}q^\alpha
                    \right)\;\;\nonumber\\
              &&\mbox{} - \mbox{} {16\over3}{p_+^2p_-^2\over a^3}
                    \left(
                    \vphantom{g^{\mu\beta}}
                    p_+^\mu q^\alpha - p_+\!\!\cdot q \;g^{\mu\alpha}
                    \right) p_+^\beta\nonumber\\
              &&\mbox{} + \mbox{} {16\over3}{p_+^2p_-^2\over a^3}
                    \left(
                    p_-^\mu q^\beta - p_-\!\!\cdot q \;g^{\mu\beta}
                    \right) q^\alpha
\end{eqnarray}
with $p_+^2>0\;,\;p_-^2>0\;,\;q^2=0$ and $a\equiv p_+^2-p_-^2$.
Note that the first term is proportional to the tree-level $WW\gamma$ vertex.
The cut diagrams corresponding to the cut $p_-^2>0$ are related by
crossing symmetry. Adding the four cut diagrams, one ends up with:
\begin{equation}
\sum_{\mbox{\tiny cuts}} Z^{\alpha\beta\mu} =
{8\over3}\;2\left(  g^{\alpha\beta}p_+^\mu
                  +g^{\mu\alpha}q^\beta
                  -g^{\mu\beta}q^\alpha
          \right)\;\;.
\label{eq:Zepfac}
\end{equation}
Inserting the overall factor and the fermion lines, one sees that the extra
diagrams amount to a scaling of the $WW\gamma$ vertex with
$1+i\Gamma_{_W}/M_{_W}$. It should be noted, that the factorization of the
correction is not universal.

However, to get electromagnetic current conservation in the process
(\ref{process1}), one can also effectively write the correction
(\ref{eq:Z}) in this form.
In the limit $q^2\to0$, the overall factor multiplying the standard
Yang-Mills vertex is then given by
\begin{equation}
\label{eq:approx}
        1 + i\frac{\gamma_{_W}(p_+^2)}{p_+^2 - p_-^2}\;\;.
\end{equation}
The parts in (\ref{eq:Z}) transverse to $q^\mu$ are dropped since they
do not play a role in restoring electromagnetic current conservation.
Only if one would allow for a negative running width in the $t$-channel,
rather than taking $\gamma_{_W}(p_-^2)=0$, multiplying the standard
Yang-Mills vertex with an overall factor $1+i\Gamma_{_W}/M_{_W}$ would give
a result that respects electromagnetic gauge invariance.

These simple, effectively factorizing prescriptions for restoring
electromagnetic gauge invariance may be easier to implement in a Monte Carlo
generator.  However, in general they violate the full SU(2)$\times$U(1) gauge
invariance and, even more, upset the balance between the diagrams taking part
in the unitarity cancellations at high energies.  Hence, the validity is
limited to the low energy range $\sqrt{s} = \ord{M_{_W}}$, e.g., LEP2.  In
contrast, the factorized form obtained from Eq.~\ref{eq:Zepfac}, being exact,
does not have this problem.

%  #] gauge cancellation 1 :
%  #[ numerical results 1 :

\section{Numerical results for $e^-e^+\to e^-\bar{\nu}_e u \bar{d}$}
\label{sec:numerics}

The process $e^-e^+\to e^-\bar{\nu}_e u \bar{d}$ has been studied numerically.
The fermions are all taken to be massless, except for the electron, which has
a mass $m_e$. The input parameters are given below,
\begin{eqnarray}
m_e         & = & 0.511\cdot10^{-3}\; \mbox{GeV}\;\;,\nonumber\\
M_{_W}         & = & 80.22\; \mbox{GeV} \;\;,\nonumber\\
\alpha(0)      & = & 1/137.036\; \;\;,\nonumber\\
G_F         & = & 1.16\cdot 10^{-5}\;\mbox{GeV}^{-2}\;\;,\nonumber\\
\sqrt{s}    & = & 175\;\mbox{GeV}\;\;,\nonumber\\
50\;\mbox{GeV} \le & \sqrt{p_+^2} & \le 110\;\mbox{GeV}\;\;.
\end{eqnarray}
The fermionic width of the $W$ boson is computed using Eq.~\ref{widthw}. This
gives $\Gamma_{_W} = 2.02773\ldots$ GeV.

The cross-section for $e^-e^+\rightarrow e^-\bar{\nu_e}u\bar{d}$ for the
different schemes for two values of the minimum electron scattering angle
$\theta_{\rm{min}}$ are given in table \ref{tab1}.

\begin{table}[htb]
\begin{center}
\begin{tabular}{|l|rr|}\hline\hline
Scheme                    & \multicolumn{2}{c|}{Cross-section [pb]} \\
                          & $\theta_{\rm{min}} = 0^\circ$ &
                                             $\theta_{\rm{min}} = 10^\circ$ \\
\hline
Fixed width                               & .08887(8)   & .01660(3) \\
Running width, no correction              & 60738(176)  & .01713(3) \\
Fudge factor, with running width          & .08892(8)   & .01671(3) \\
Pole scheme, with running width           & .08921(8)   & .01666(3) \\
Running width, with Eq.~\ref{eq:Z}        & .08896(8)   & .01661(3) \\
Running width, with Eq.~\ref{eq:approx}   & .08897(8)   & .01662(3) \\
\hline\hline
\end{tabular}
\end{center}
\caption[]{Total cross-section for
$e^-e^+\rightarrow e^-\bar{\nu_e}u\bar{d}$ in different schemes.}
\label{tab1}
\end{table}

\begin{figure}
\begin{center}
\unitlength=1mm
\begin{picture}(140,140)
\epsfxsize=140mm
\put(0,0){\mbox{\epsfbox{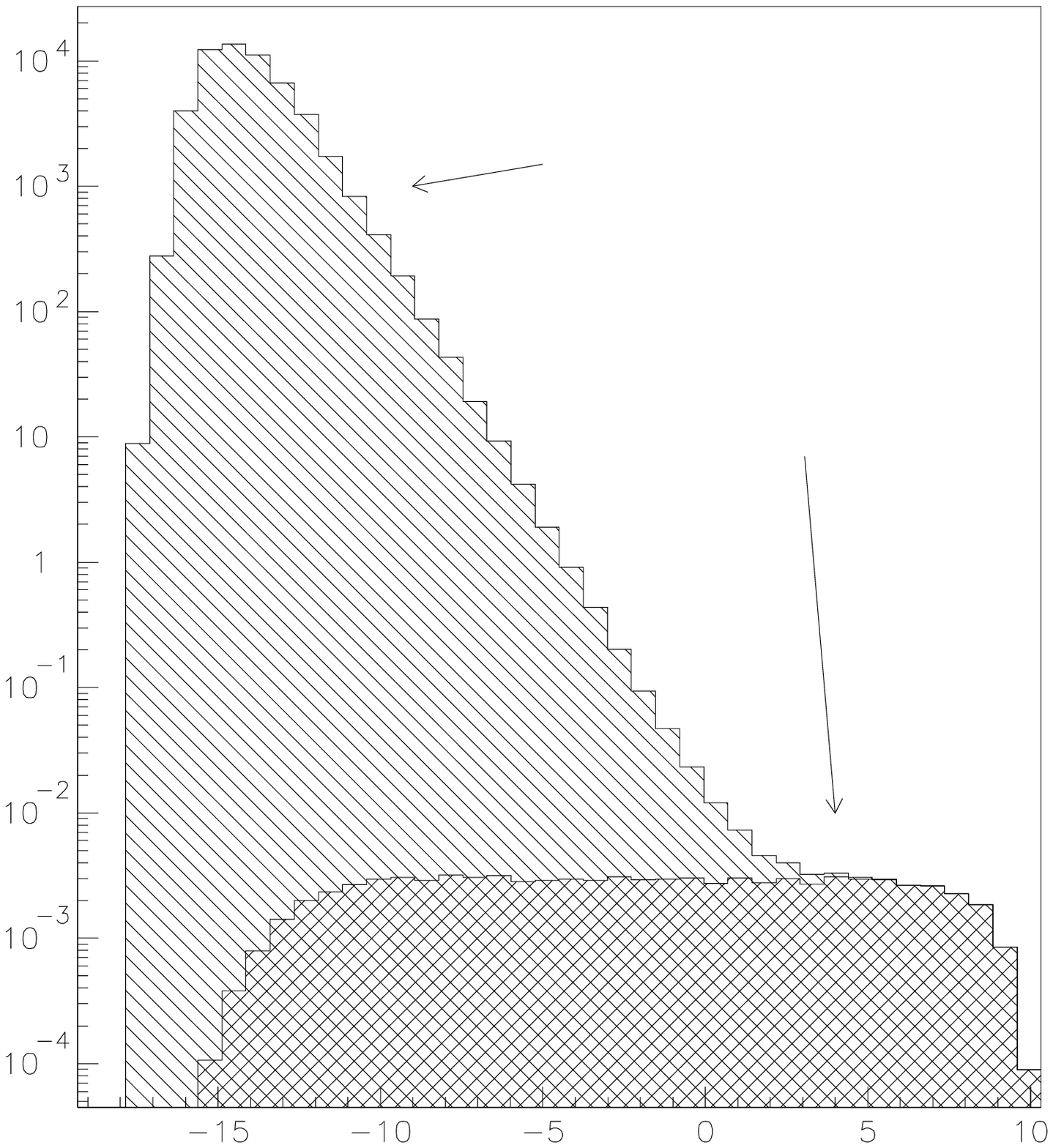}}}
\put(127,1){\makebox(0,0)[br]{\large$\ln(-q^2/\mathrm{GeV}^2)$}}
\put(5,127){\makebox(0,0)[tl]{\large\shortstack{$\displaystyle q^2
    \frac{d\sigma}{dq^2}$\\[1pt][pb]}}}
\put(76,106){\shortstack{Running width,\\no correction}}
\put( 90,82){\shortstack{Running width,\\with correction}}
\end{picture}
\end{center}
\caption[]{The effect of gauge-breaking terms in $e^-e^+\to
e^-\bar{\nu}_e u \bar{d}$ as a function of the virtuality of the photon.}
\label{results_eeenuud}
\end{figure}

Note that all schemes were computed using the same
sample, so the differences are much more significant than the integration
error suggests.  One sees that in this case, once current conservation is
restored the results for the total cross-section of the different methods
agree to $\ord{\Gamma_{_W}^2/M_{_W}^2}$.
{}From Fig.~\ref{results_eeenuud} it should be clear, that if we include
running-width effects without taking into account the correction of the
Yang-Mills vertex, too many events are sampled for small values of $q^2$.

%  #] numerical results 1 :
%  #[ conclusion:

\section{Conclusion}

The violations of gauge invariance associated with a naive introduction of a
finite width for unstable particles can have disastrous consequences.  We
have indicated that, in the case of the vector bosons, this can be cured in
a fully consistent way by the inclusion of appropriate fermionic corrections,
e.g., to the three-vector-boson vertex.  It
has been shown explicitly in the case of massless fermions and the $WW\gamma$
vertex that the electromagnetic Ward identity is restored and current
conservation holds.  In the process $e^-e^+ \to e^- \bar{\nu}_e u \bar{d}$,
in which gauge-breaking terms are amplified by $\ord{10^{11}}$, this is
shown to lead to a correct result.  The differences between this scheme and
other ways to obtain a gauge-invariant result have been shown to be small,
much less than $\Gamma_{_W}/M_{_W}$ in this specific example.
The correction to the $WW\gamma$ vertex is
given explicitly in Eq.~\ref{eq:Z} for current conserving sources, and in a
simplified factorized form suitable for this process at not too high energies
in Eq.~\ref{eq:approx}.  These functions can be incorporated in other event
generators for LEP2.

%  #] conclusion:
%  #[ acknowledgements :

\section*{Acknowledgments}
This research has been partly supported by the EU under contract
numbers CHRX-CT-92-0004 and CHRX-CT-93-0319.
J.~Hoogland and G.~J. van Oldenborgh are supported by FOM.
G.~Passarino is supported by INFN.
We would also like to thank the CERN Liquid Support Division.

%  #] acknowledgements :

%  #[ bibliography :

%\bibliographystyle{prsty}
%\bibliography{fenomeen}

%  #] bibliography :

%  #[ figures :

%  #] figures :
\end{document}